\documentclass[journal]{ieee_style}

\usepackage{amsfonts}
\usepackage{graphicx}
\usepackage{amssymb}
\usepackage{amsmath}
\usepackage{latexsym}
\usepackage{caption}
\usepackage{mathtools}
\usepackage{url}
\usepackage{array}
\usepackage{hyperref}
\usepackage{float}

\begin{document}
\title{Power-Based Side-Channel Attack for AES Key Extraction on the ATMega328 Microcontroller}

\author{Utsav Banerjee,
        Lisa Ho,
        and Skanda Koppula
\thanks{All authors are with the Department
of Electrical and Computer Engineering, Massachusetts Institute of Technology, Cambridge,
MA, 02139 USA}
\thanks{To contact the authors: \texttt{utsav@mit.edu}, \texttt{lisaho@mit.edu}, and \texttt{skandak@mit.edu}}
\thanks{Manuscript completed for 6.858 Computer Systems Security; completed on December 5, 2015}}

\markboth{6.858 Final Project Report - Fall 2015}%
{Power-Based Side-Channel Attack for AES Key Extraction on the ATMega328 Microcontroller}
\maketitle

\begin{abstract}
    We demonstrate the extraction of an AES secret key from flash memory on the ATMega328 microcontroller (the microcontroller used on the popular Arduino Genuino/Uno board). We loaded a standard AVR-architecture AES-128 implementation onto the chip and encrypted randomly chosen plaintexts with several different keys. We measured the chip's power consumption during encryption, correlated observed power consumption with the expected power consumption of the plaintexts with every possible key, and ultimately extracted the 128-bit key used during AES. We describe here our test infrastructure for automated power trace collection, an overview of our correlation attack, sanitization of the traces and stumbling blocks encountered during data collection and analysis, and results of our attack.
\end{abstract}


\section{Introduction}
Recent concerns about data privacy have brought attention to encryption algorithms. One of the more popular symmetric-key algorithms, Advanced Encryption Standard (AES), has been the U.S. government standard since 2002 (ISO/IEC 18033-3), and is used in a multitude of applications: SSL/TLS protocols \cite{ssl}, Kerberos \cite{kerberos}, and demonstrably secure embedded devices \cite{embedded}. This last application in particular, embedded devices, has seen much growth in recent years, given the advantages of computation on smaller embedded devices: low power, lower system latency, and generally smaller device size.

Small hardware implementations, however, are notoriously vulnerable to a range of side-channel attacks \cite{smalldevice}. Timing, electromagnetic radiation, and power consumption are just three commonly exploited vectors used to leak information about ongoing computations and data on the chip. Knowing that a device architecture is vulnerable to side-channel exploitation is useful in deciding whether to execute unprotected sensitive computations or store data on devices with similar memory and processor characteristics.

We aim to demonstrate a reasonably realistic power-based side-channel attack on AES-128-ECB software implementation on one such embedded device: the ATMega328 microcontroller produced by Atmel. The ATMega328 is the basis for the widely popular development board, Arduino Uno \footnote{Other models of the Arduino, such as the Arduino Mega and Arduino Genuino Micro use ATMega chips as well, that have a similar architecture to the ATMega328. It is possible that this attack could be adapted to those chips as well.}.

In section II, we review the theoretical ideas underpinning our attack. In section III, we describe our experiment: our hardware setup, power measurement infrastructure, correlation methods, instructive problems that we encountered, and overview of the structure of our source code. In section IV, we quantitatively describe the results of our attack.

\section{Preliminaries}
\subsection{Controller Specifications}
The ATMega328 family of chips is an 8-bit microcontroller series with 32 KB of NAND-type flash and 2KB of SRAM. The controller runs off a 16 MHz external clock on the Arduino board. Typical power consumption of the chip is a 20mA current draw from 5V power supply, but it can vary depending upon the peripheral and I/O pin usage \cite{atmeldatasheet}. Our attack exploits the NAND-type flash memory architecture that consumes marginally more power when accessing addresses that store value-zero (discharge) bits \footnote{In a highly simplified power consumption model, NAND-flash charges a central bit line connected to a series of memory cells. Depending on the value stored in the accessed memory cell, the line is discharged or not. Thus, the line requires data-dependent recharging.} \cite{nandflash}.

The encryption program running on our ATMega328, \texttt{AESLib}, uses an Arduino-specific port of the \texttt{avr-crypto-lib} by Davy Landman and Bochum Hackerspace \cite{AESLib} \cite{daslabor}. \texttt{AESLib} is one of the more widely-used AES implementations for Arduino, and includes support for ECB and CBC-modes of AES. Our team decided that ECB-mode would be more vulnerable to a power correlation attack, and correspondingly chose to exploit the library's AES-ECB implementation. We discuss ECB in further depth in section IIID.

\subsection{Correlation Power Analysis}
Correlation Power Analysis (CPA) is a type of side channel attack that relies on power consumption information. On a high level, CPA attempts to correlate observed power consumption with expected power consumption. To a greater extent than more basic forms of power analysis such as Simple Power Analysis, CPA attacks are able to extract secret keys from noisy data \cite{cpa}. This requires collecting the power consumption of a device performing encryption over many different plaintexts with the same key. We can then build a power model that contains the expected power consumption of the device performing a particular operation during encryption over the given plaintexts with every possible key. 

In our case, we use CPA against AES encryption. Several operations during AES are good candidates to model power consumption for. Attacks tend to model power from the first round of AES because it provides the best data regarding which bytes of plaintext and key are combined. Power analysis tends to target the XOR operation in the \texttt{AddRoundKey} step or the SBOX substitution during the \texttt{SubBytes} step, since these can consume predictable amounts of power. Different CPA implementations use different power consumption models; the most common are the Hamming Weight of the operation result or the Hamming Distance of the input and output of the targeted operation. Hamming Weight and Hamming Distance may be correlated with the amount of power consumed because reading a ``0'' or a ``1'' from memory may require different amounts of power.
 
After deciding on a power model to use, using CPA against AES requires collecting the amount of power consumed over encryption of hundreds of plaintexts, and then calculating the expected power consumption for each of these plaintexts with every possible key. Trying every possible key with AES is possible because a different byte of the key is used for every byte of the plaintext. This reduces our search space to $2^8$ possibilities for each key byte. 

After building a power hypothesis consisting of the amount of power consumed for each possible plaintext with every possible key byte, we can calculate the correlation coefficients between these expected amounts of power and the amount of power observed during encryption. One advantage of CPA is that we don't need to know exactly when our targeted operation occurs during encryption; we can calculate the correlation coefficient between the power hypothesis and the power trace (collected power consumption over time) at each point of the trace. We then take the key byte that gives us the maximum correlation coefficient as our best guess key byte. Repeating this process 16 times gives us best guesses for all 16 bytes of the key.

\section{Protocols and Procedure}
\subsection{Data Collection Infrastructure}
\begin{figure}[ht]
\centering
\includegraphics[width=2.5in]{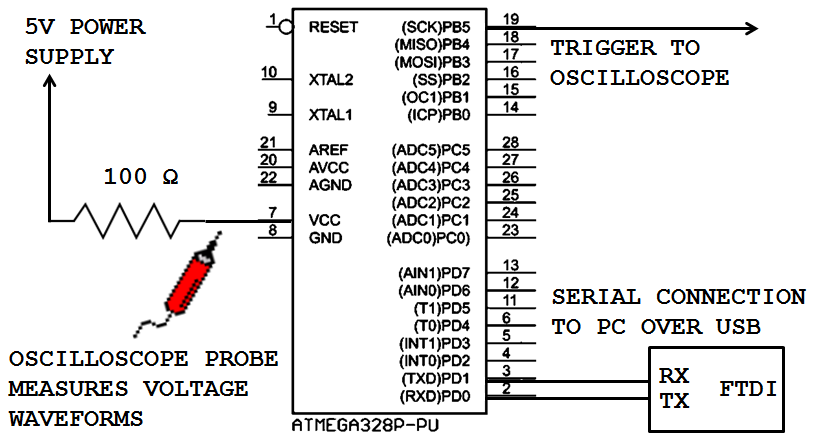}
\caption{The oscilloscope probes voltages on the chip's power line, starting automated power trace on a trigger that signals the start of AES}
\label{fig_sim}
\end{figure}

A central piece of our work was developing the serial-connection based data collection framework to feed plaintexts for encryption to the Arduino and read the resulting power trace from the oscilloscope. An outline is given in Fig. \ref{fig_sim}.

We needed a subset of the power trace that correlated strongly with the input plaintext and key. Specifically, we were interested in finding the portion of the trace that corresponded to the \texttt{xor(plaintext,key)} operation, and the the SBOX permutation taken over the \texttt{xor} result. In order to automate slicing to this part of the trace, we modified the \texttt{AESLib} SBOX assembly code to insert a flag on a memory-mapped register that pulled up pin 13 on the board (pin 19 on the chip). Our oscilloscope triggered on this output and automatically captured the part of the trace immediately after the rising edge on pin 13.

\begin{figure}[!t]
\centering
\includegraphics[width=2.5in]{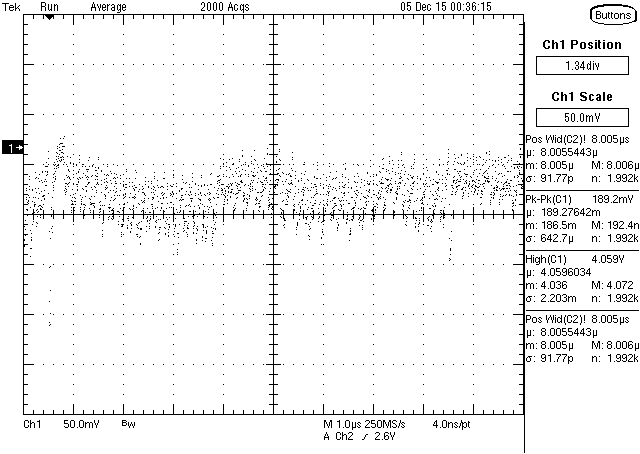}
\caption{Voltage measured during AES encryption, as shown on the oscilloscope}
\label{pow}
\end{figure}

Central to our data collection was a Tektronix 5054B computing oscilloscope that collected samples at a maximum of 4GHz with internal memory bank of at most 16 million points. The sampling rate limited the resolution of traces that we were able to capture; we originally thought this physical cap on data quality was causing correlation issues we ran into, until we discovered other issues related to DC shifts and time shifts which we discuss in more detail in Section III.C. An example of a typical power trace we collected is shown in Fig. \ref{pow}. It may look like the sampling rate is lower than desirable, but we explain this more in Section III.C.

We used the oscilloscope's GPIB (IEEE-488 General Purpose Interface Bus) query interface to automate configuration and downloading of traces.

In the final iteration of our system, we have an orchestrating computer $C$ send plaintexts to the Arduino for encryption every 2 seconds over the Arduino's serial port (Fig. \ref{fig_sim}). The pin 13 trigger resulted in an oscilloscope trace capture, which was subsequently sent back to $C$. 

More recently, we have been attempting to use the Tektronix $Fast Frames$ feature to take capture batches of 2,500-point traces at once, and use one memory-read and GPIB-write operation to transfer these traces to the computer. This would allow us much faster trace measurement, which is currently bottlenecked by the GPIB-write operation. As we will discuss in Section IV, the number of key bytes we can recover is directly correlated with the number of plaintexts we can capture. 

Photos of our collection framework can be found at \texttt{\url{https://www.dropbox.com/sh/usialgelvfqlsdr/AACvqOHKEWoumWNYi2WRIebCa?dl=0}}. 

\subsection{Implementation of CPA and Power Model}

We examined \texttt{AESLib}, our AES implementation of choice, to try to determine how to estimate power consumption most effectively. The XOR operation from the \texttt{AddRoundKey} step consisted of two operations, a load and an XOR: 1) \texttt{ld r0, X+}, and 2) \texttt{eor param, r0}, where \texttt{param} was a byte of the plaintext and \texttt{X+} was a byte of the key. The SBOX operation from the \texttt{SubBytes} step consisted of a move and load program memory instruction: 1) \texttt{mov r30, ST00}, and 2) \texttt{lpm ST00, Z}. We hypothesized that reading from memory might require different amounts of power depending on whether we were reading a ``0'' or a ``1''. 

We first modelled the power consumed during the SBOX operation, since its nonlinearity would make for clearer key results. However, when we built a power model based on the SBOX operation and then tried to correlate it to our first set of traces, we found no useful correlation. It later became clear that this was a problem with our data rather than our power model (as outlined in III.C). At the time, however, we were concerned that maybe the power consumed by the Arduino was for some reason not proportional to the SBOX output. 

\begin{figure}[!t]
\centering
\includegraphics[width=2.5in]{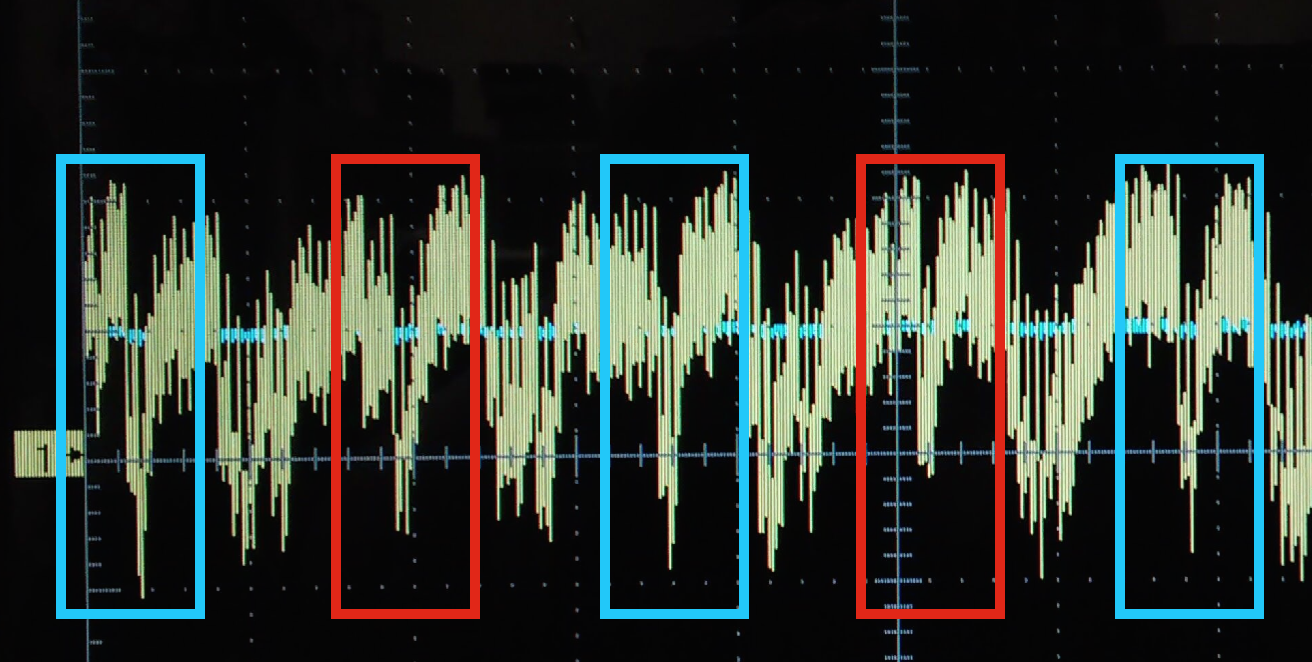}
\caption{XOR operations alternated between results of all "0"s and all "1"s. The power consumption was noticeably different depending on the Hamming Weight of the result.}
\label{fig_xor}
\end{figure}

To verify that the results of the SBOX and XOR operations were actually correlated with the amount of power the Arduino consumed,  we examined the power consumption of just the AVR assembly operations corresponding to these steps atomically. We were unable to easily make out any difference in power consumption between SBOX operations that resulted in all ``0''s or all ``1''s, but we were able to easily see a diffference in power consumption between XOR operation that resulted in all ``0''s or all ``1''s (Fig. \ref{fig_xor}).

Based on these results, we moved forward with correlating the measured power consumption with the Hamming Weight of the XOR result. After collecting more reliable traces (as explained in III.C), we re-ran CPA using this metric and were able to recover many of the key bytes. However, we ran into a symmetry issue. Each key byte and its complement with respect to 255 had an equal correlation with the power consumption (e.g. key byte candidates such as 1 and 255 or 3 and 252 showed the same correlation magnitude). Taking this into account, we were able to correctly identify $\frac{11}{15}$ key bytes down to the correct key byte \texttt{k} or \texttt{255-k}. 

This symmetry from the XOR output meant we wouldn't be able to recover the correct key without applying brute force guessing after running CPA. Using the output of the SBOX for the power model would have the advantage of avoiding this symmetry, giving us one correct key. After getting more reliable traces, we retried CPA with a power model employing the Hamming Weight of the SBOX output and found that this was actually a reliable power model to use. Using Hamming Weight of SBOX output while increasing the number and reliability of traces we used, we were able to recover all key bytes.

 In later iterations of our code, we aimed to speed up analysis of waveforms and obtain results like those in Fig. \ref{results_random2} more quickly. To do this, we replaced our correlation function with a faster iterative correlation function that calculates the correlations given $1,2,3,\ldots, i, \ldots, num\_plaintexts$ traces, without recalculating correlations for the previous $i-1$ traces.

\subsection{CPA Challenges and Solutions}
During the course of the experiment, we faced several challenges in data collection and interpretation. They are listed below, along with our solutions:
\begin{itemize}
\item Our initial idea was to measure the voltages at the two ends of a $1\Omega$ resistor connected between the VCC pin of the microcontroller and the power supply, and measure their difference using a Tektronix ADA400A Differential Pre-amplifier. However, as with any practical amplifier, the ADA400A has a gain-bandwidth product limitation, and we could achieve a gain of 10 at our required bandwidth. Also, the amplifier introduced its own noise into the measured power traces, which reduced our chances of a successful CPA attack. To resolve this issue, we finally decided to get rid of the differential measurement setup, and measure only the voltage at the VCC pin. We also replaced the $1\Omega$ resistor with a $100\Omega$ one, in order to get better gain.
\item Another challenge in CPA data collection was dealing with DC shifts in the power waveform. Minor fluctuations in the power supply voltage are natural, but they can mask or scramble the power spikes due to the AES operations. Our solution was to measure power traces for 10 iterations of AES encryption on the same plaintext and take the average of these 10 waveforms.  The averaging operation was done using the oscilloscope, and it indeed provided significant improvements in our attack potency. 
\item We noticed that some of our AES computations took different lengths of time than others, even though they were supposedly doing the same operations. We suspected that these differences in computation duration may have been due to interrupts. To test this hypothesis, we disabled interrupts during AES computation by clearing the Global Interrupt flag. This greatly reduced the timing differences from trace to trace, improving trace alignment during correlation.
\item We also faced the issue of missing traces. When the Arduino encrypted some number of plaintexts, the oscilloscope would record the power consumption from most, but not all, computations. This made it difficult for us to match power traces with the correct plaintext. We realized that the oscilloscope was missing traces because it did not always have time to reset itself between different plaintexts. Adding a delay before sending the next plaintext to the Arduino gave the oscilloscope enough time to reset and correctly record power traces from all of our plaintexts.
\item Some of our traces were quite noisy. To minimize the amount of noise in our samples, we measured the encryption of each plaintext ten times and took the average. We also added a grounded aluminum box to our setup, which may have reduced electromagnetic noise and interference.
\item We originally used the maximum sample rate while collecting data (4 GHz), which meant that our data took a long time to collect and download. To speed up this process, we reduced our sample rate to 2500 MHz. Even at this lower sample rate, we were able to recover all necessary information with at most 600 plaintexts. It might be interesting to further examine the relationship between sample rate and number of plaintexts required in order to find optimal combinations.
\item CPA essentially relies on statistical algorithms. Therefore, it is extremely important to have large sets of data samples to analyse. We started our experiment with only 100 power traces (100 random plaintexts) and were able to reliably recover only 10-11 of the 16 key bytes. Soon, we realized that we needed more data, and repeating the experiment for 300-400 traces (plaintexts) enabled us to recover all the key bytes. It has to be noted that some keys needed even more power traces, and we estimate that 1000 traces are enough to recover any key. Further data and graphs from our experimental results can be found in the link mentioned in Section IV.
\end{itemize}

\subsection{Limitations}
Aspects of our attack implementation are unlikely to be possible in real-world attack scenarios; we address these concerns and suggest possible remedies:
\begin{itemize}
    \item[--] We chose to attack the Arduino library's implementation of the AES-ECB mode, rather than the more secure CBC-mode. However, ECB is still (unfortunately) used as the default option in a number of crypto-suites, \texttt{avr-crypto-lib} included, despite it not being semantically secure (i.e. you can derive information about the plaintext from the ciphertext). This is because of its relatively simple implementation, compared to other more sophisticated modes of AES. Furthermore, our attack does not exploit the plaintext-ciphertext correlations in ECB to derive the key; rather, it uses the power sidechannel. For our team's first power-trace based attack, we chose a mode that we were confident might have some correlation with the plaintext-key XOR (the first computation in the first step of ECB); however, it might be possible to adapt our attack to CBC-mode as well.
    \item[--] We modified the \texttt{AESLib} assembly code to pull high one of Arduino's I/O to signal the oscilloscope when to begin collecting traces (the scope's trigger). In a real world attack, it is unlikely that we will be able to insert a convenient signal to aid our trace collection and analysis result. Nonetheless, careful manual analysis of one trace could yield a constant time frame of interest, which, barring waveform time shifts, we could use to automate trace collection and trace slicing. We have demonstrated ways to reduce time-shift errors in our work.
\end{itemize}

\subsection{Overview of Source Code}
All our source code can be found at \texttt{\url{https://github.com/skoppula/aes-sidechannel}}. Files in the source that might be of interest:
\begin{itemize}
    \item[--] \texttt{cpa/cpa.py} contains the source that reads in the waveform binary data (with helper \texttt{wfm2read\_fast.py}) and subsequently runs CPA. It outputs the top ten key byte guesses per byte of key, and a correlation visualizations like Fig. \ref{results_random2}.
    \item[--] \texttt{data-capture/arduino-aes-2} runs AES encryption on a new plaintext 10 times, each time a `run' command is received on the terminal. The ten traces are later averaged on the oscilloscope. \texttt{data-capture/arduino-aes-1} runs AES encryption on a plaintext sent over the serial port.
    \item[--] \texttt{data-capture/scope-interface-2} runs a Python script that interfaces with the Arduino to run AES at regular intervals, and interfaces with the oscilloscope to collect the corresponding traces. \texttt{data-capture/scope-interface-1} is a Matlab implementation of similar functionality that we originally used to interface with the oscilloscope over GPIB; interfacing with the Arduino was done seperately in this version of the code (in \texttt{data-capture/send-plaintexts-processing})
    \item[--] \texttt{data-capture/process-wfms} contains the scripts we put together initially to parse and plot the first \texttt{*.wfm} waveform binaries that we recieved from the oscilloscope. This was later integrated into \texttt{cpa/cpa.py}
\end{itemize}

\section{Results}
We conducted the experiment three times, with three different secret keys: two randomly generated keys and one non-random key \footnote{By non-random, we mean the key \texttt{[0x0 0x1 0x2 0x3 0x4 0x5 0x6 0x7 0x8 0x9 0xA 0xB 0xC 0xD 0xE 0xF]}. This was the test key chosen during the development stage of our project, as we were building the trace-collection infrastructure and writing the CPA code.}. In all three, we were able to discover the correct key.

\begin{figure*}[]
\centering
\includegraphics[width=7in]{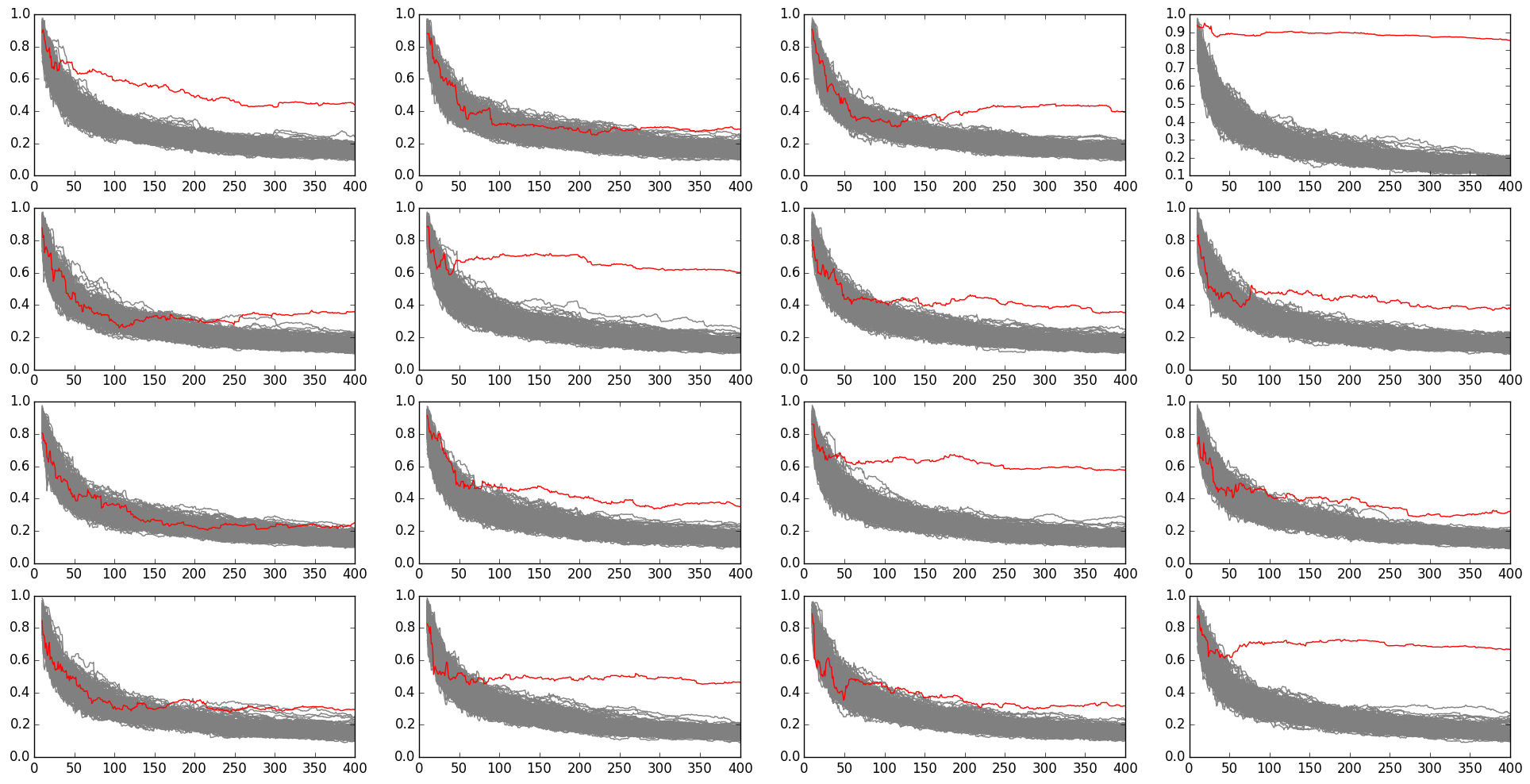}
\caption{Each graph corresponds to one of the 16 blocks/16 plaintext bytes/16 key bytes in our AES implementation. A single graph plots the correlation of all 256 key guesses (y-axis) with the power trace data, as more data is fed into the correlation (x-axis). The red line represents the correct key guess. Notice that by the time we feed the 400th trace into our CPA implementation, the correct key byte is the key guess with the highest correlation with all the traces (visible in the height of each red line at $x=400$). Results for the other two CPA runs can be found at the Dropbox link.}
\label{results_random2}
\end{figure*}

However, interestingly, the number of traces required to recover all key bytes differed from key to key. Specifically, it took 600 trace averages (each corresponding to a plaintext) to get complete accuracy on the non-random key, 300 trace averages to completely uncover the first random key, and 400 trace averages to uncover the second random key. We were interested in understanding the performance of each key guess as the number of input plaintext traces increased; for each run, we created a plot that tracked each key guess's correlation with the trace data as the amount of trace data increased (Fig. \ref{results_random2})

It takes roughly 30 minutes for our infrastructure to collect the traces for the encryption of 100 plaintexts. Data collection time seems to scale linearly with the number of plaintexts we process: encrypting and collecting the traces for 500 traces takes on average two and half hours. Processing these 500 traces and extracting the correlation values and key estimates over $i$ plaintexts for all $i$ (Fig. \ref{results_random2}) takes roughly five to ten minutes.

Further data and results can be found at \texttt{\url{https://www.dropbox.com/sh/07xni6s4tu4klme/AABnrBK-QZCVO1tMK4GFeQ5ta?dl=0}}.

\section*{Acknowledgments}
We would like to extend our deepest thanks to Chiraag Juvekar of the Energy-Efficient Circuits and Systems group for the time he spent with us aiding our debugging of data collection and analysis problems. We would also like to thank Albert Kwon, our TA, for his insightful advice over the course of the project.

\end{document}